# Predict the water level of the Lake Mead for the next 30 years based on ARIMA

Yixin Li

Beijing 101 High School

Abstract

In this study, a mathematical model is developed for the drought problem of Lake Mead. First, a polynomial fitting of the elevation of Lake Mead to the area of the lake is done by the least-squares method, and the volume of Lake Mead is approximated by the numerical integration of the product of the height and the area solved by the trapezoidal rule. The accuracy of the fitting reached more than 96% at all four different locations. Second, the minimum and maximum water levels were transformed into volume numbers by the above method, and the historical data of Lake Mead were classified into three classes of water resources by sequential clustering. According to these data, the optimal cut point of the most recent drought period was 2008 and has continued until now. Finally, two prediction models were constructed using ARIMA(2,2,2) and ARIMA(3,2,2) to study the water level data from 2008 to 2020 and 2005 to 2020, respectively, to predict the water level data of Lake Mead from 2022 to 2050, and to compare and analyze them.

Keywords: trapezoidal rule, ordered clustering, ARIMA

## 1. Introduction

Lake Mead is a reservoir on the border of Arizona and Nevada and is the largest reservoir in the United States. As a result of the ongoing drought and rising demand, Lake Mead's water level is reported to be at its lowest ever in the summer of 2021, dropping to 36 percent of its full capacity. At the same time, the number of states suffering from water scarcity due to drought is rising, and researchers are focusing on the Lake Mead drought problem in order to solve the problem. Elliot A. Alexander et al [1] studied how to cope with the water demand of Lake Mead by using a multi-objective evolutionary algorithm (MOEA) in combination with other models; Ranjan Parajuli et al [2] developed the LVV model to predict the future water demand of Lake Mead; Samuel Potteiger et al [3] studied the water demand of Lake Mead by using the UA model to study Lake Mead inflow. In this paper, we will develop a mathematical model based on the Lake Mead drought problem, our contributions are as follows:

- We study the relationship between the elevation, area, and volume data of Lake Mead

- We study the water level change of Lake Mead under different conditions by ordered clustering and then divide the drought periods.
- We compare the long-term water level change trend of Lake Mead from 2005 to 2020 and the short-term water level change trend in the dry period through the ARIMA model.

## 2. Relationship of elevation, area, and volume

The shape and area of Lake Mead's surface change as the water level rises and falls[4]. This work employs the technique of least squares to establish a polynomial fit in order to investigate the relationship between elevation, area, and volume. The volume of Lake Mead is approximated through numerical integration of the product of height and area after first studying the relationship between elevation and area.

### *2.1. Least Squares Solving Polynomial Functions of Elevation and Area*

The least-squares method uses the sum of squares of the residuals to find the best function match for the data[5]. We fit the elevation and area data $P_i(x_i, y_i)(i = 1,2,3,4)$ provided by the Lake Mead official website as a cubic function in this study:

$$f(x_i) = \theta_0 + \theta_1 x_i + \theta_2 x_i^2 + \theta_3 x_i^3 \qquad (1)$$

where $x_i$ is the elevation of Lake Mead, $y_i$ is the area of Lake Mead, and $\theta_j (j = 1,2,3,4)$ is the coefficient of the polynomial. The sum of squares of errors for each data point within the sample data for the elevation and area of Lake Mead is:

$$S = \sum_{i=1}^{4} [f(x_i) - y_i]^2 \qquad (2)$$

We use the maximum likelihood estimation method to solve for each coefficient of the polynomial $\theta_j (j = 1,2,3,4)$ such that the error sum of squares S obtains a minimal value. Thus, for the optimal function, the partial derivatives of the error sum of squares S with respect to each polynomial coefficient $\theta_j (j = 1,2,3)$ satisfy[6]:

$$\frac{\partial S}{\partial \theta_j} = \sum_{i=1}^{4} \left[ 2(\theta_0 + \theta_1 x_i + \theta_2 x_i^2 + \theta_3 x_i^3 - y_i) x_i^j \right] = 0 \qquad (3)$$

The relationship between Lake Mead elevation and area is obtained by using MATLAB as follows:

$$f(x_i) = -3113260.7431 + 9079.6851 x_i - 8.9000 x_i^2 + 0.0029 x_i^3 \qquad (4)$$

The fitting result is shown in the following figure 1:

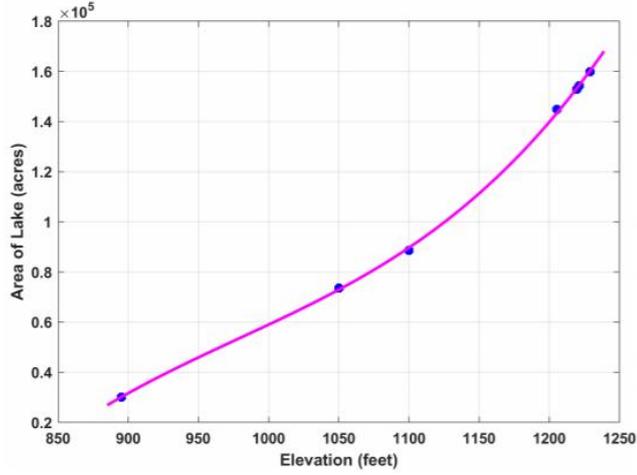

Figure 1: 3 multi-fold minimum two-squares fitting

## 2.2. Relationship of elevation, area, and volume

To further solve the relationship between the elevation, area, and volume of Lake Mead, this paper uses the ideas from the infinitesimal method. The minimum to maximum values of the elevation is infinitesimal by combining the fitting results of the previous elevation and area. We assume that the discrete distance of the elevation is $\Delta x$, and multiply the elevation and its area for each $\Delta x$ distance to obtain the volume of the lake for each $\Delta x$ elevation, which can be expressed as the following equation in terms of a continuous equation:

$$V = \int_{x_{min}}^{x} f(x) \approx \sum_{i=1}^{n}(V_i) = V_0 + \cdots + V_n \tag{5}$$

where $x_i$ is the elevation at $i \times \Delta x$ feet and $V_i$ is the volume of the lake at $i\Delta x$ feet. To solve the above integral values, the trapezoidal integration method is used in this paper. If the integral of $f(x)$ over $[a, b]$ is to be solved, the integration interval needs to be divided into $n$ segments with equal length[7]. As a result, the distance between every two segments is $\Delta x = \frac{b-a}{n}$, and then the approximate solution is carried out as the following figure 2. Thus, the integral value over this interval, i.e., the volume of Lake Mead, is approximately equal to the sum of the areas of each small trapezoid.

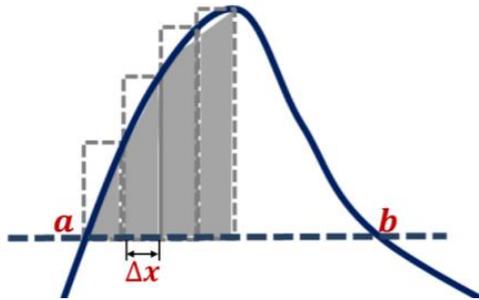

Figure 2: Schematic diagram of trapezoidal rule integration

The upper base $f(x_0)$ and the lower base $f(x_1)$ of the first trapezoid (leftmost) at the first $\Delta x$ are $\Delta x = \frac{b-a}{n}$[7], so the corresponding Lake Mead The volume is as follows:

$$V_1 = \big(f(x_0) + f(x_1)\big) * \Delta x / 2 = \frac{\big(f(x_0) + f(x_1)\big) * \Delta x}{2} \tag{6}$$

And so on, the upper base $f(x_{n-1})$, the lower base $f(x_n)$ of the elevation at the nth $\Delta x$, the height is $\Delta x = \frac{b-a}{n}$, so the corresponding volumes of Lake Mead are as follows:

$$V_1 = (f(x_{n-1}) + f(x_n)) * \Delta x / 2 = \frac{(f(x_{n-1}) + f(x_{n1})) * \Delta x}{2} \qquad (7)$$

Therefore, the volume of Lake Mead at the elevation $n \times \Delta x$ can be obtained by the trapezoidal rule as follows:

$$V = \frac{(b-a)}{2*n}\left[f(x_0) + 2\sum_{i=1}^{n} f(x_i) + f(x_n)\right] \qquad (8)$$

The relative error of the fit is defined as follows:

$$\text{Accuracy} = \left(1 - \frac{|V_{\text{observe}} - V_{\text{predict}}|}{V_{\text{observe}}}\right) \times 100\% \qquad (9)$$

where $V_{\text{observe}}$ is the observed value of the volume, and $V_{\text{predict}}$ is the fitted value of the volume. In this paper, the volumes obtained by polynomial fitting and trapezoidal rule integral are compared with the volume corresponding to the four elevations (1229.0, 1219.6, 1050.0, 895.0 feet) provided by the official website, as shown in the following table 1:

Table 1 Quantitative table of indicators

| Elevation(feet) | Volume observations (acre-feet) | Volume fit (acre-feet) | Accuracy |
|---|---|---|---|
| 1229.0 | 29686054 | 30239879 | 0.008 |
| 1219.6 | 28229730 | 28868386 | 0.022 |
| 1050.0 | 10217399 | 10716135 | 0.048 |
| 895.0 | 2576395 | 2576395 | 0 |

The relationship between the elevation, area, and volume of the Lake Mead is obtained through the above method, as shown in the following figure 3:

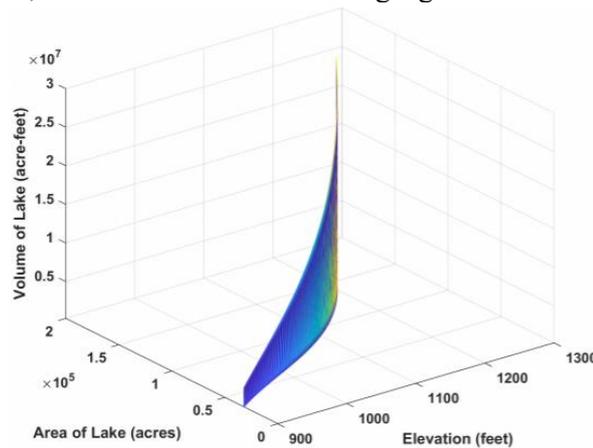

Figure 3: 3D graph of Lake Mead volume changing with elevation and area

## 3. Divide the arid period by sequential clustering

The water levels of Lake Mead are different in different periods. Based on the historical data of the highest and lowest water levels of Lake Mead from 1935 to 2020, it can be seen that the water levels of Lake Mead vary greatly in different periods. In order to study the drought trend of Lake Mead, this paper divides the situation of Lake Mead in different periods into three levels, namely adequate levels, moderate levels, and drought levels. Through the model for solving the volume in Part 2, the annual minimum lake volume and maximum lake volume corresponding to the minimum water level of Lake Mead from 1935 to 2020 are obtained. Then, the minimum and maximum lake volumes of Lake Mead from 1935 to 2020 were sorted in descending order, represented by the set $f = SD_1, SD_2, \cdots, SD_{66}$. It is divided into three categories, and there are $C_{m-1}^{N-1}(m = 3, N = 66)$ cutting methods[8]. Calculate the best cut point by sequential clustering, the steps are as follows:

**Step1**: define the class diameter:

$$D(i,j) = \sum_{i=1}^{j} |X_t - \widetilde{X}_G| \quad (10)$$

where $\widetilde{X}_G$ is the median of the lake volume data for Lake Mead.

**Step2**: Define the loss function for classification

Use $b(n,k)$ to represent a certain classification of dividing n ordered samples into k categories. In the model of this paper, the historical water level of Lake Mead is divided into three states, so this paper takes $n = 66, k = 3$, and the loss function is as follows:

$$L[b(66,3)] = \sum_{j=1}^{3} D(j_t, j_{t+1} - 1) \quad (11)$$

**Step3**: Solve for the optimal solution

To solve the optimal classification method $P(n,k)$ we need to minimize the smallest loss function. We first find the point $j_3$ to make $L[P(n,3)] = L[P(j_3 - 1, 2) + D(j, 66)]$ is the smallest, so the third class $G_3\{j_3, j_3 + 1, \cdots, 66\}$ [9], then find $j_2$ so that it satisfies the following formula:

$$L[P(j_2, 2)] = L[P(j_2 - 1, 2) + D(j_2, j_3 - 1)] \quad (12)$$

Thus, we obtain the second class $G_2\{j_2, \cdots, j_3 - 1\}$, and by using similar methods all classes $G_1, G_2, G_3$ can be obtained in turn, which is the optimal solution we need, namely $P(66,3) = \{G_1, G_2, G_3\}$.

Through the above steps, this paper divides the minimum lake volume and the maximum lake volume of Lake Mead from 1935 to 2020 respectively. The first critical point for classifying the drought class according to the lowest volume class of Lake Mead is 15527229acre-feet, and the second critical point is 20862371acre-feet, as determined by ordered clustering. The first critical point to classify the drought according to the highest volume class of Lake Mead is 16038974acre-feet and the second critical point is 23354267acre-feet. The resulting visualization of the calculated annual minimum and maximum volumes and the division of drought are shown in figure 4:

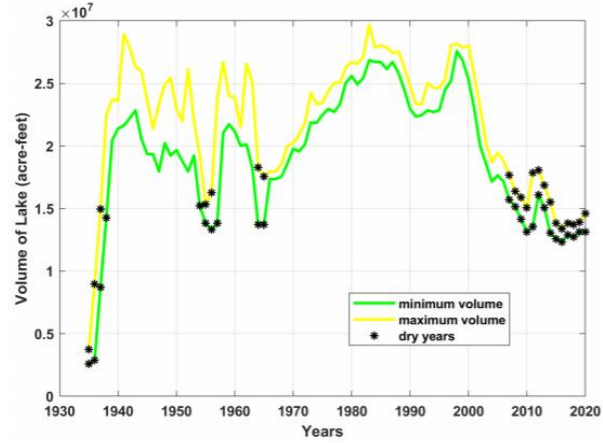

Figure 4： Results of drought division

## 4. Predict the water level of Lake Mead by using the ARIMA model

Given the low stability of the historical water level data of Lake Mead, and considering the actual water level and water volume will be affected by regional consumption, wastewater recycling, climate temperature, precipitation, etc., there will be a certain change trend. Therefore, this paper needs to consider the dependence between the sequence, combined with the consideration of the time sequence model. As a result, we use the differential autoregressive moving average model (ARIMA) to predict and analyze the water level data of Lake Mead. The ARIMA model combines the autoregressive model, the moving average model, and the difference method. In ARIMA, AR is autoregression, MA is moving average, p is the autoregressive term, q is the order of the moving average, and d is the order of difference that needs to be performed on the data[10].

**Basic algorithm idea:** In this paper, the historical water level data $W_t$ of Lake Mead is subjected to $d = 0,1,\cdots,n$ times difference processing to obtain a new stable data series $W_t$ of Lake Mead water level. Then the ARIMA(p,q) model is fitted by $W_t$ and the original d-time difference is restored. Therefore, the predicted data of the future Lake Mead water level can be obtained. The algorithm steps are as follows:

**Step1：** Differential processing of time series

The difference operation is performed on the historical data of the water level of Lake Mead, and the water level data is recorded as $X_t$. In this paper, the first-order difference operation is performed, as shown below:

$$\nabla X_t = X_t - X_{t-1} = W_t \qquad (13)$$

then we have

$$W_t - 0.5 W_{t-1} = \varepsilon_t \qquad (14)$$

$\nabla$ is called 1st-order backward difference operation, $W_t$ is the sequence of Lake Mead water level data after a first-order difference operation, after such a first-order difference operation. The original non-stationary sequence $X_t$ is transformed into a stationary sequence $W_t$. The general expression for ARIMA(p,q) is:

$$W_t = \varphi_1 W_{t-1} + \varphi_2 W_{t-2} \cdots + \varphi_p W_{t-p} + \epsilon_t - \theta_t \epsilon_{t-1} - \cdots - \theta_p \epsilon_{t-p}, t \in Z \qquad (15)$$

In the formula, the first half is the autoregressive part, the non-negative integer p is the autoregressive order, $\varphi_1,\cdots,\varphi_p$ are the autoregressive coefficients. the second half is the moving average part, and the non-negative integer q is the moving average order, $\theta_1,\cdots,\theta_p$ are the moving average coefficients; $W_t$ is the data sequence of the historical water level of Lake Mead, and $\epsilon_t$ is $WN(0,\sigma^2)$.

The stationary preprocessing is done on the sequence, and a coherence test is done on the sequence $W_t$ after each time of difference until the data obtained from the difference can pass the stationarity test. The times of difference are denoted as d-time difference and obtain a new stationary sequence $W_1, W_2, \cdots, W_{t-d}$.

**Step2**：Zero mean processing

We let all Lake Mead water level data as observation data, and subtract the average value of this group from each a group of data to obtain a new group of data, namely: $W'_t = W_t - \overline{W}$. As a result, we get a set of preprocessed newness series $W'_t$.

**Step3**：ACF & PACF test

We calculate the autocorrelation function (ACF) $\rho_k$ and partial autocorrelation function (PACF) $\varphi_{kk}$ of the sequence $W'_t$ after preprocessing the historical water level data of Lake Mead, and determining the model that $X'_t$ fits.

**Step4**：Parameter estimation and model ordering

The AIC and BIC criteria are used to determine the model order, that is, the parameters of the overall ARIMA model are estimated through the sample distribution of the historical data of Lake Mead water level.

### 4.1. model1: Only data from the most recent dry period is considered

Based on the calculation in Part 3, the most recent arid period is from 2008 to 2020. Through the analysis of the historical water level data of Lake Mead from 2008 to 2020, we calculate that the autoregressive term p=2 in Model 1, the order of the moving average q=2, and the order of difference d=2; the fitting result using the prediction model 1 is shown in the following figure 5:

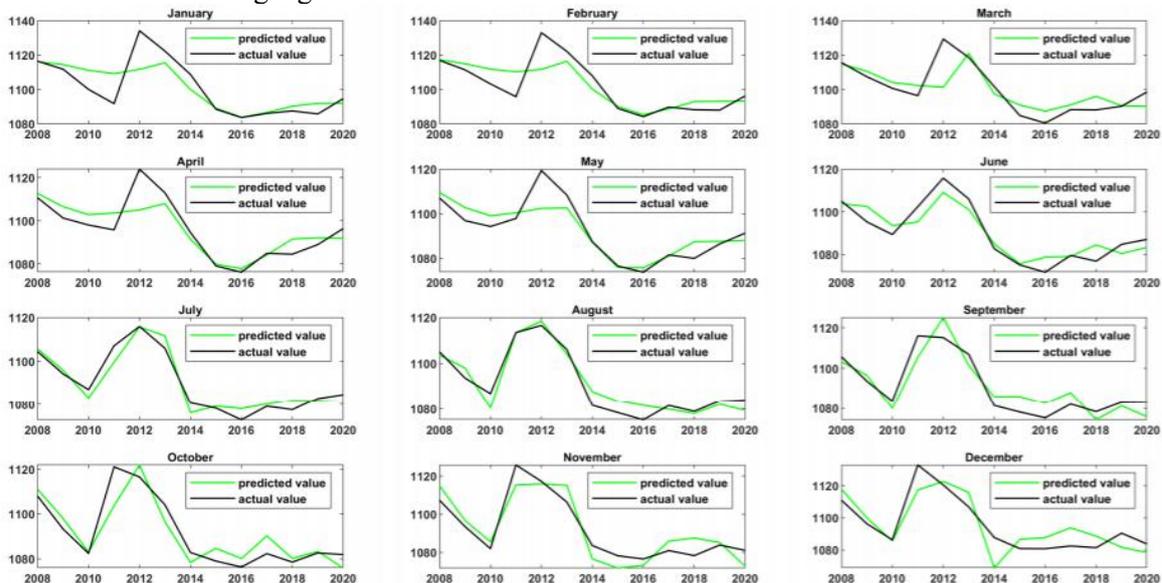

Figure 5:The water level of Lake Mead from January to December from 2021 to 2050 is predicted by Model 1

### 4.2. model2: Consider data from 2005 to 2020

Similarly, based on the analysis of the historical water level data of Lake Mead from 2005 to 2020, we calculate that in the ARIMA model, the autoregressive term p=3, the order of the moving average q=5, and the order of difference d=2; The fitting results using prediction model 2 are shown in the following figure 6:

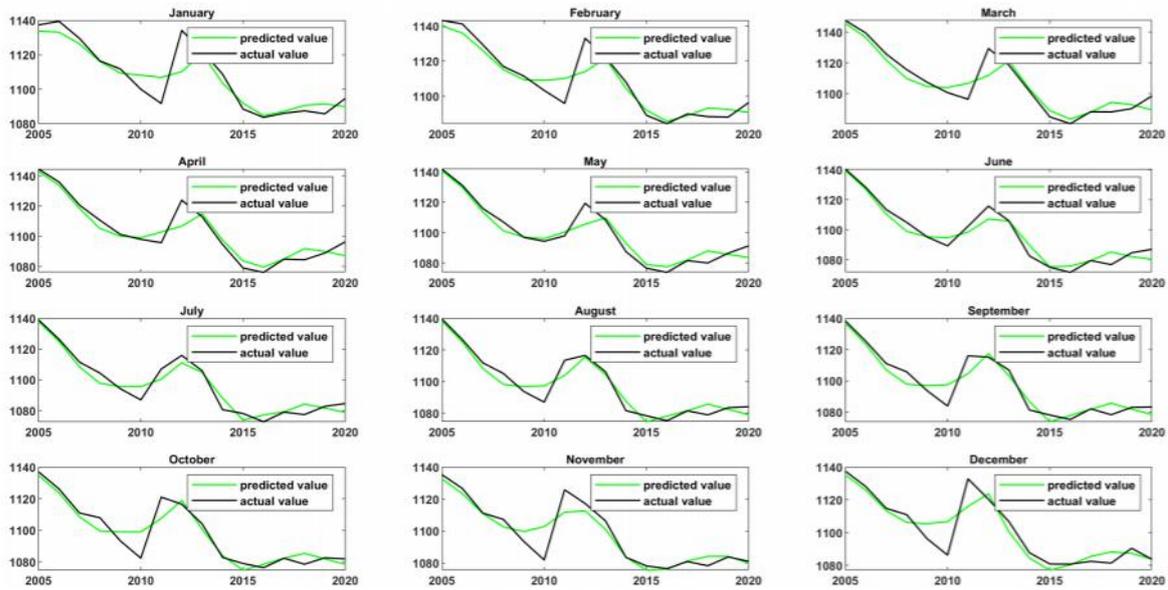

Figure 6: The water level of Lake Mead predicted by Model 2 from January to December 2021 to 2050

### 4.3. Comparison and Analysis of Models 1 and 2

The ARIMA(2,1,2) model is used in Model 1 and the ARIMA(3,1,2) model is used in Model 2 to predict the water level changes of Lake Mead from January to December from 2021 to 2050. The comparison chart is as follows shown:

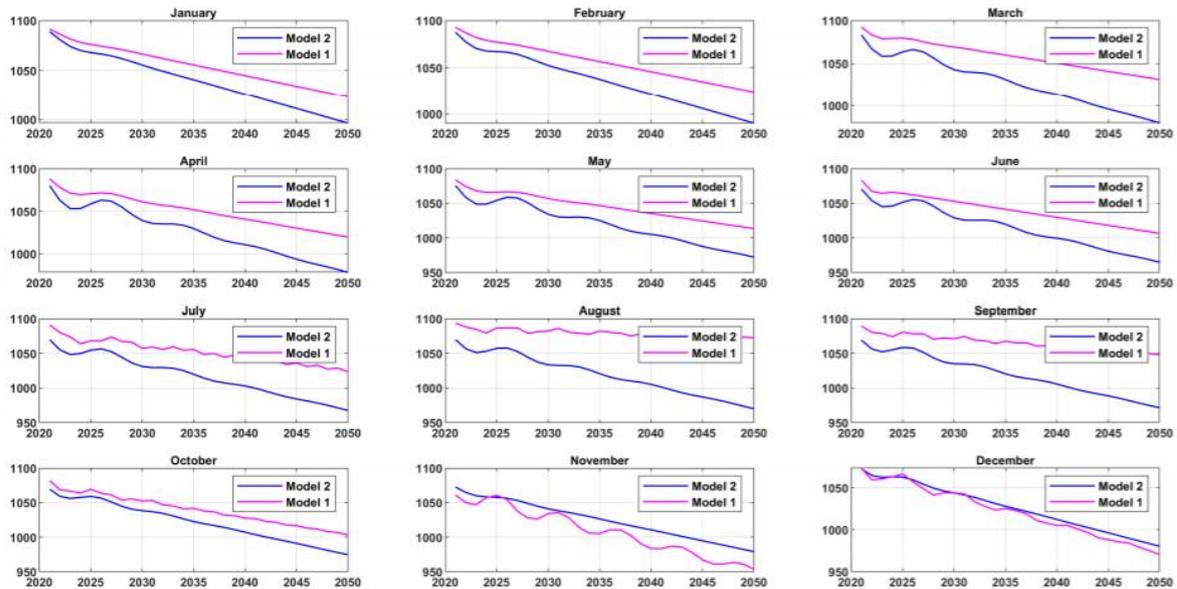

Figure 7: Lake Mead water levels predicted by Model 1 and Model 2 from January to December

As can be seen from the above figure 7, the water level of Lake Mead predicted by Model 2 in the next 30 years is generally lower than the water level predicted by Model 1. At the same time, the decline was even greater. By observing the original data, it can be seen that the data from 2005 to 2020 is used in the model whereas only the data from 2008 to 2020 is used in model 1. According to the original data, from 2005-to 2007, both the minimum water level and the maximum water level of Lake Mead showed a continuous downward trend. The lowest water level of Lake Mead decreased from 1130.01(feet) in 2005 to 1104.41(feet) in 2008, and the highest water level decreased from 1147.8(feet) in 2005 to 1117.96(feet) in 2008.

It can be seen that compared with Model 1, Model 2 includes a period of historical data that the water level has dropped significantly. In the process of adaptive fitting, the ARIMA model predicts the future change trend according to the changing trend of its existing data, so the predicted results obtained by Model 2 are more rapid decline than those of Model 1, and it can also be seen that the water level of Lake Mead has decreased significantly in recent years.

## 5. Conclusion

In this paper, we use the least square method to fit the polynomial of the height and area of Lake Mead. The volume is approximated by integrating the height of Lake Mead and the corresponding area by the trapezoidal rule.

Secondly, we use the ordered clustering to divide the historical water level of Lake Mead into three stages, and the critical point of the dry period was obtained.

Finally, in order to study the long-term and recent trends of the water level of Lake Mead, the ARIMA model was used to study two models: 1. The recent drought data obtained by ordered clustering, that is, the water level data of Lake Mead from 2008 to 2020; 2. Lake Mead's water level data does not include the drought period, that is, the water level data of Lake Mead from 2005 to 2020. And based on these two models, we predict the water level changes of Lake Mead from 2022 to 2050. Since the minimum and maximum water levels of Lake Mead have continued to decline in these three years, the prediction result obtained by Model 2 is declining faster than that of Model 1. It can also be seen that in recent years, the water level of Lake Mead has dropped significantly.